# Lite-FBCN: Lightweight Fast Bilinear Convolutional Network for Brain Disease Classification from MRI Image

Dewinda Julianensi Rumala[1], Reza Fuad Rachmadi[1,2], Anggraini Dwi Sensusiati[3], I Ketut Eddy Purnama[1,2]

[1]Department of Electrical Engineering, Institut Teknologi Sepuluh Nopember, Surabaya, Indonesia
[2]Department of Computer Engineering, Institut Teknologi Sepuluh Nopember, Surabaya, Indonesia
[3]Department of Radiology, Universitas Airlangga, Surabaya, Indonesia
Correspondence Author: ketut@te.its.ac.id

**Abstract**
Achieving high accuracy with computational efficiency in brain disease classification from Magnetic Resonance Imaging (MRI) scans is challenging, particularly when both coarse and fine-grained distinctions are crucial. Current deep learning methods often struggle to balance accuracy with computational demands. We propose Lite-FBCN, a novel Lightweight Fast Bilinear Convolutional Network designed to address this issue. Unlike traditional dual-network bilinear models, Lite-FBCN utilizes a single-network architecture, significantly reducing computational load. Lite-FBCN leverages lightweight, pre-trained CNNs fine-tuned to extract relevant features and incorporates a channel reducer layer before bilinear pooling, minimizing feature map dimensionality and resulting in a compact bilinear vector. Extensive evaluations on cross-validation and hold-out data demonstrate that Lite-FBCN not only surpasses baseline CNNs but also outperforms existing bilinear models. Lite-FBCN with MobileNetV1 attains 98.10% accuracy in cross-validation and 69.37% on hold-out data (a 3% improvement over the baseline). UMAP visualizations further confirm its effectiveness in distinguishing closely related brain disease classes. Moreover, its optimal trade-off between performance and computational efficiency positions Lite-FBCN as a promising solution for enhancing diagnostic capabilities in resource-constrained and or real-time clinical environments.

## 1. INTRODUCTION

Brain disease recognition is a pivotal aspect of clinical diagnosis and treatment planning, where timely and accurate identification of conditions can significantly impact patient outcomes [1]. Magnetic Resonance Imaging (MRI) serves as a non-invasive imaging modality that provides detailed views of brain structures, making it an essential tool in detecting and monitoring brain diseases [2]. However, interpreting these complex images requires advanced computational techniques to assist radiologists and clinicians in making precise diagnoses.

In recent years, Deep Learning techniques, particularly Convolutional Neural Networks (CNNs), have emerged as powerful tools for automated analysis of medical images, including MRI scans [3], [4]. CNNs have shown remarkable success in various medical imaging tasks, including brain disease recognition from MR images. Despite their effectiveness, CNNs typically require large annotated datasets to generalize well to new, unseen data. This dependency can be problematic in clinical settings where data may be scarce, diverse, and costly to obtain. Furthermore, the standard CNN architectures often struggle to capture complex feature interactions, which are crucial for distinguishing subtle variations in medical images.

To address the limitations of conventional CNNs, researchers have developed Bilinear CNNs (BCNNs) that enhance the model's ability to capture higher-order feature interactions by combining feature vectors from two separate networks through bilinear pooling method [5]. BCNNs have demonstrated improved performance in various image recognition tasks [6]–[9]. However, the dual-network design of BCNNs results in high computational complexity and significant memory usage [10], making them less feasible for deployment in resource-constrained environments such as outpatient clinics, rural healthcare facilities, and other settings with limited computational resources typically found in clinical practice.

In response to these challenges, we propose Lite-FBCN, a Lightweight Fast Bilinear Convolutional Network tailored for brain disease recognition from MRI images. Lite-FBCN addresses the computational inefficiencies of conventional BCNNs by incorporating a single lightweight CNN backbone for feature extraction, thereby reducing the overall computational cost and memory footprint. Additionally, Lite-FBCN introduces a channel reducer before the bilinear pooling layer, which compresses the feature map dimensionality and results in a more compact bilinear vector. This design does not only decrease the model size but also accelerates the inference time.

Moreover, the complexity and subtlety of brain MRI features necessitate an advanced method capable of capturing fine-grained details, even when applied to inter-class brain condition recognition tasks such as distinguishing between cognitive normal (CN), neoplastic (NEO), cerebrovascular (CVA), degenerative disease (DGD), and inflammatory/infectious disease (INF). Conventional CNNs might struggle with these subtle distinctions due to their limited ability to model higher-order feature interactions. Lite-FBCN, by leveraging the bilinear pooling mechanism in a more efficient manner, enhances the recognition of these subtle patterns and interactions within MRI scans. This is crucial for accurate diagnosis and effective treatment planning.

## 2. RELATED WORKS

Researchers have proposed CNN-based models for a wide range of medical imaging tasks, such as lesion detection, organ segmentation, and disease classification [1], [11]–[14]. These studies have demonstrated the potential of CNNs to achieve state-of-the-art performance in diagnosing various medical conditions from different imaging modalities, including MRI, CT, and PET. However, the growing demand for efficient and deployable medical image analysis systems has led to the development of lightweight CNN architectures. These models are designed to balance computational efficiency and classification performance, making them well-suited for resource-constrained environments. Examples of lightweight CNN architectures include MobileNets [15], [16] and EfficientNet [17], which achieve compact network structures without compromising accuracy.

Bilinear CNNs [5], which model pairwise feature interactions using bilinear pooling, have shown promising results in various computer vision tasks, including fine-grained image classification and scene understanding [6]–[9]. In the context of medical image analysis, researchers have explored the application of bilinear CNNs, especially for disease diagnosis using from various imaging modalities [18]–[21]. These studies have demonstrated the ability of bilinear CNNs to capture spatial relationships and discriminative features essential for accurate diagnosis from medical images.

However, BCNNs require high computational resource due to the utilization of two different backbone networks. To address this issue, Fast Bilinear Convolutional Networks (Fast BCNNs) have been proposed [20] to reduce computational complexity. In Fast BCNNs, bilinear interactions are derived from the single backbone network, which contrasts with the dual-network design of conventional BCNNs. Fast BCNNs has been applied to various image classification tasks, such as for remote sensing image scene classification [7] and breast cancer classification [20], demonstrating its effectiveness in reducing computational overhead while preserving model performance.

In the case of brain disease recognition from MRI images, several studies have investigated the use of CNNs for this task by focusing not only on binary classification [2], [22]–[24], but also multi-class brain disease classification [2], [25]–[30]. Brain diseases often exhibit complex and subtle features that are challenging to distinguish, necessitating advanced feature extraction methods. Thus, different methods have also been proposed for this task, including transfer learning [24], [27], [30]–[33], ensemble learning [11], [27], [32], [34], [35], and even BCNNs [10]. While CNNs with various enhancedment-based approaches have shown promising results in these tasks, there remains a need for lightweight and efficient models capable of accurately classifying brain diseases in resource-constrained clinical settings.

## 3. ORIGINALITY

This paper introduces Lite-FBCN, a novel Lightweight Fast Bilinear Convolutional Network specifically designed to address the computational and performance limitations of conventional Bilinear CNNs (BCNNs) for brain disease recognition from MRI scans. The originality of Lite-FBCN lies in its innovative approach to combining high accuracy with significantly reduced computational complexity. Unlike traditional BCNNs that rely on bilinear interaction from dual-network architectures, Lite-FBCN implements a single-network design where the captured information from this backbone alone is processed through the bilinear pooling layer.

A key design aspect of Lite-FBCN is the introduction of a channel reducer before the bilinear pooling layer to reduce the dimensionality of the backbone's feature maps, thus lowering the computational complexity of the bilinear pooling operation. This results in a smaller, more efficient model with reduced inference time that maintains high accuracy in brain disease recognition. The channel reducer ensures that the feature maps are condensed without losing critical information, thereby enhancing the model's ability to capture intricate feature interactions necessary for distinguishing subtle variations in MRI images.

The versatility of Lite-FBCN is evident in its ability to achieve high accuracy with lightweight backbone networks. This adaptability allows Lite-FBCN to be tailored to different deployment scenarios, accommodating a range of resource constraints without compromising performance. The combination of high accuracy, reduced computational complexity, and flexibility in backbone selection underscores the originality and potential impact of Lite-FBCN in advancing the state of the art in brain disease recognition from MRI images. This makes Lite-FBCN a promising tool for enhancing diagnostic capabilities in clinical practice.

## 4. SYSTEM DESIGN
### 4.1 Proposed Lite-FBCN

The proposed Lite-FBCN (Lightweight Fast Bilinear Convolutional Network) aims to enhance brain disease recognition from MRI scans by addressing the computational and performance limitations of traditional Bilinear Convolutional Neural Networks (BCNNs). Lite-FBCN introduces a streamlined, single-network design that leverages transfer learning for the feature extraction backbone. This approach significantly reduces the computational cost compared to conventional BCNNs, which rely on dual-network architectures. Additionally, Lite-FBCN incorporates a channel reducer layer before the bilinear pooling layer, compressing the feature map dimensionality and further decreasing computational complexity. This combination of techniques not only maintains high accuracy but also ensures faster inference times, making Lite-FBCN suitable for deployment in both resource-constrained and real-time clinical environments. The schematic of the proposed Lite-FBCN design is illustrated in Figure 1.

In this research, we focus on a brain disease classification task involving the differentiation between several categories: cognitive normal (CN), neoplastic (NEO), cerebrovascular (CVA), degenerative disease (DGD), and inflammatory/infectious disease (INF). This classification task is critical for providing accurate and timely diagnosis, which can impact patient outcomes.

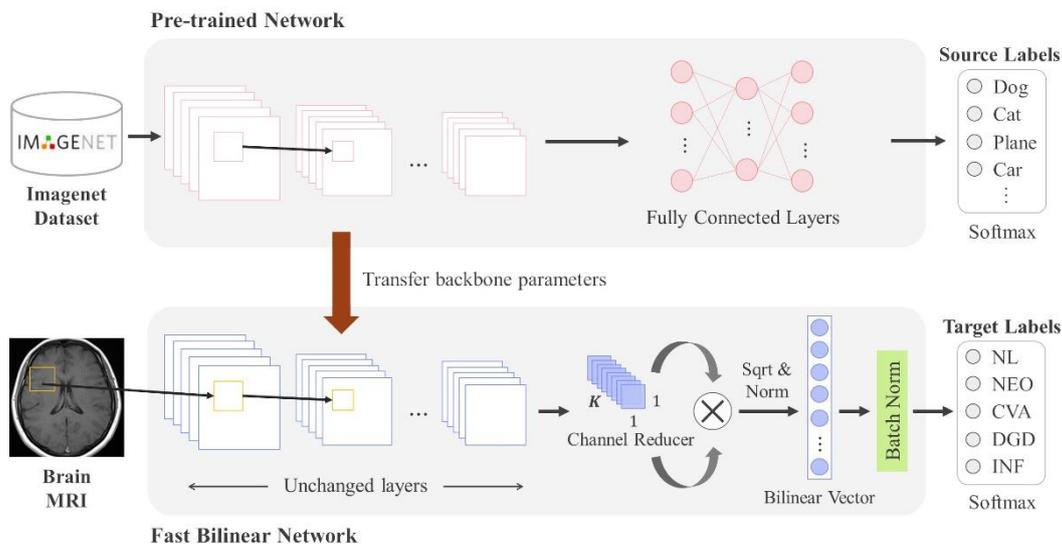

**Figure 1.** Schematic diagram of the proposed Lite-Fast Bilinear Convolutional Network (Lite-FBCN).

### 4.1.1 Backbone Networks

The models used in this study are based on the well-established lightweight pre-trained CNN models, including MobileNetV1 [15], MobileNetV2 [16], and EfficientNetB0 [17]. We employ these CNN models as the backbone network in this study due to its capability to extract low-level features using its smaller kernel. To enhance the performance of the backbone networks and mitigate distributinoal data problem, transfer learning is applied in this work. The data source for the transfer learning application that we used in this study is natural images, derived from ImageNet. During the training, we fine-tuned all layers of the CNN models, so that all weights are optimized according to the learned features starting from the low-level to the high-level, as there are high distinctions between the natural images and medical images [1], [36].

### 4.1.3 Channel Reducer Layer

The channel reducer layer, placed before the bilinear pooling layer, is implemented as 1×1 convolutional layer, which effectively combines the information across the input channels and reduces their number according to the specified number of filters. This layer is particularly applied to reduce the feature map output of the backbone network prior to engaging in bilinear interactions.

Let $F \in \mathbb{R}^{H \times W \times C}$ be the feature map output of the backbone network, where $H$ is the height, $W$ is the width, and $C$ is the depth or the number of channels. The channel reducer layer, implemented as 1x1 convolutional layer with $K$ filters will reshape this the number of channels of the original feature maps $F$. If $W_{cr} \in \mathbb{R}^{1 \times 1 \times W \times K}$ represents the weights of the channel reducer layer, then the output feature map $\boldsymbol{F_{cr}} \in \mathbb{R}^{H \times W \times K}$ of the channel reducer layer can be computed according to equation 1.

$$\boldsymbol{F_{cr}} = \boldsymbol{Conv_{1 \times 1}}(F, W_{cr}) \qquad (1)$$

The number of channels in $\boldsymbol{F}$ is resized into $\boldsymbol{F_{cr}}$ according to the number of channel reduction factor (CRF), denoted as $\boldsymbol{\gamma}$, which is defined as the quotient of the original number of channels $\boldsymbol{C}$ divided by the number of filters $\boldsymbol{K}$ used in the channel reducer layer as expressed in equation 2.

$$\boldsymbol{\gamma} = \frac{C}{K} \qquad (2)$$

The channel reducer layer plays a crucial role in making the bilinear pooling method practical and effective. By reducing the dimensionality of the last feature maps $\boldsymbol{F}$ of the backbone network into $\boldsymbol{F_{cr}}$, the channel reducer lowers computational and memory costs, focuses on the most relevant features, helps in preventing overfitting, and facilitates a more efficient and stable training process. Applying this layer before the bilinear pooling layer helps the subsequent interactions captured are both manageable and meaningful, ultimately leading to better model performance.

**4.1.3 Bilinear Pooling**

Unlike standard BCNNs that utilize dual-network architectures, and Fast BCNNs that apply bilinear pooling on a single network without a channel reducer, Lite-FBCN incorporates a channel reducer to streamline the feature maps before bilinear pooling. This integration significantly reduces the computational load while preserving the model's capacity to capture detailed feature interactions. The key difference between standard Bilinear Pooling (BCNNs), Fast Bilinear Pooling (FBCNNs), and the proposed Lite-FBCN is illustrated in Figure 2.

Bilinear pooling captures interactions between all pairs of features, allowing the model to learn complex feature representations. In Lite-FBCN, we compute the bilinear features by performing an outer product of the reduced feature map $\boldsymbol{F_{cr}}$ with itself at each spatial location, followed by summation across all spatial locations. The bilinear feature $\boldsymbol{B} \in \mathbb{R}^{K \times K}$ is computed according to equation 3, where $F(i,j)$ represents a $K$-dimensional vector corresponding to the feature at spatial location $(i,j)$, and $\otimes$ denotes the outer product.

$$\boldsymbol{B} = \sum_{i=1}^{H} \sum_{j=1}^{W} \boldsymbol{F_{cr}}(i,j) \otimes \boldsymbol{F_{cr}}(i,j)^{\top} \qquad (3)$$

To improve the stability and performance of the bilinear features, normalization techniques such as signed square root and $\ell_2$ normalization is applied. After that, we apply batch normalization to enhance the stability and convergence of the network. Batch normalization normalizes the activations of each layer across the mini-batch, which helps mitigate issues such as internal covariate shift and accelerates training. Following batch normalization, the processed features are fed into a classification layer, where softmax activation is applied to produce class probabilities. To regularize the model and prevent overfitting, we apply $\ell_2$ regularization with a regularization strength of 0.01 to the weights of the classification layer.

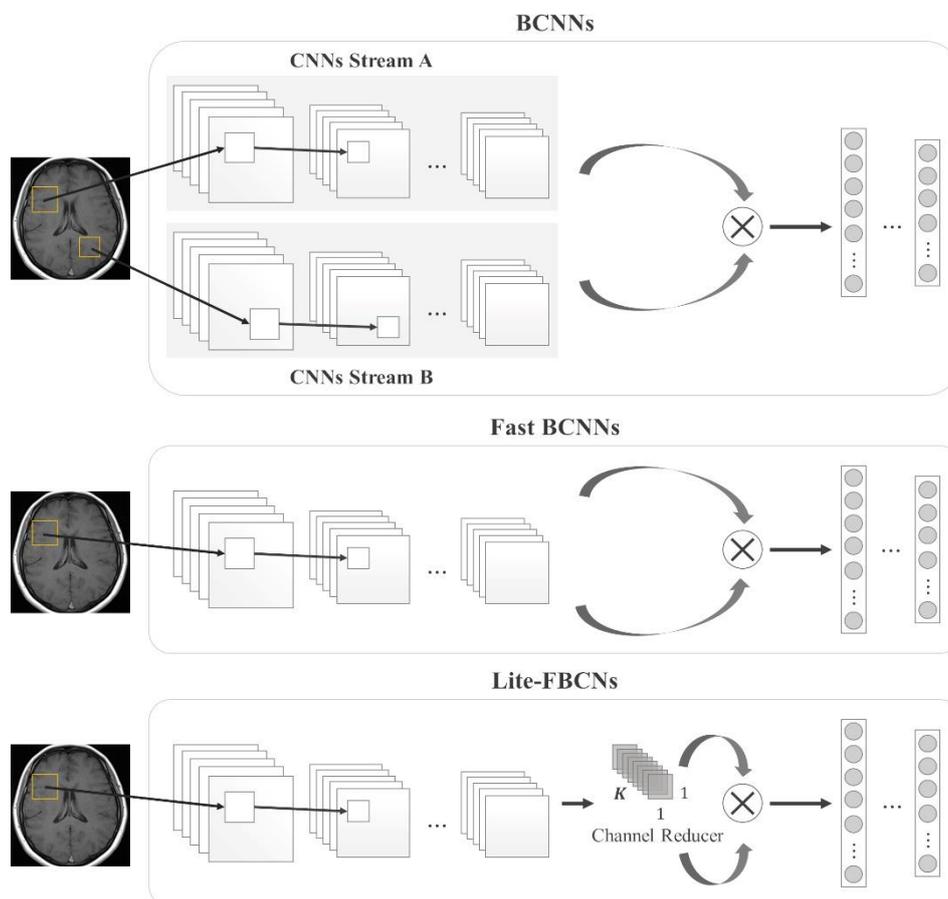

**Figure 2.** The structure differences between the standard BCNNs, Fast BCNNs, and the proposed Lite-Fast Bilinear Pooling.

## 4.2 Implementation

In this study, we utilized the dataset of T2-weighted brain MR images previously employed in a related paper concerning brain disease diagnosis [27]. The dataset underwent the same preprocessing steps as outlined in the aforementioned paper, ensuring consistency in data preparation. Furthermore, our proposed models are evaluated through five-fold evaluations for training, validation, and test sets as in the aforementioned paper. Notably, the dataset comprises multiple classes for brain disease classification, including cognitive normal (CN), neoplastic (NEO), cerebrovascular (CVA), degenerative disease (DGD), and inflammatory/infectious disease (INF), with corresponding labels consistent with those specified in the aforementioned study.

To enhance the model's robustness, we incorporated real-time data augmentation during training, introducing variability to the training datasets as outlined in [10]. We optimized the model using the Stochastic Gradient Descent (SGD) optimizer with the following parameters: a learning rate of 0.01, momentum of 0.5, and Nesterov momentum enabled. If the validation loss did not decrease for 50 epochs, the learning rate was reduced by a factor of 10, with a minimum learning rate of 0.0001. The models were trained for 500 epochs using categorical cross-entropy as the loss function. To ensure the best-performing model was retained during training, we implemented a custom model checkpoint callback as proposed in [27].

For a comprehensive insights into the methodologies, supplementary materials, and code supporting the study findings, please refer to https://djrumala.github.io/sup/lite-fbcn.

### 4.3 Performance Metrics and Evaluation

To assess the classifier model's performance, we computed accuracy, precision-recall, and F1 score. Additionally, we conducted a statistical test, repeated-measures ANOVA, to compare the proposed Lite-FCBN, the baseline, and the existing bilinear CNN models based on accuracy, precision-recall, and F1 score. A significance level of $P < 0.05$ was employed to determine statistical significance.

## 5. EXPERIMENT AND ANALYSIS
### 5.1. Performance of Baseline Models

To establish a reference for evaluating our proposed Lite-FBCN model, we first assessed the performance of several baseline models. These models include standard pre-trained CNNs that have been fine-tuned using brain MRI data for multi-class brain disease classification. Each baseline model was fine-tuned and evaluated on the same 5-fold cross-validation brain MRI data. The average performance of baseline models on the 5-fold cross-validation data is outlined in Table 1. Meanwhile, detailed results for each model's metrics per fold are available in the supplemental table. From Table 1, MobileNetV1 exhibits the highest accuracy and overall performance among these models. However, it is essential to note that our analysis revealed no statistically significant difference ($P = 0.9665$) in accuracy among the baseline models. This suggests that the choice of CNN backbone for the baseline methods may not significantly impact performance.

**Table 1.** Performance of baseline models on the 5-fold cross-validation data

| Backbone | Acc (%) | Prec (%) | Rec (%) | F1-score (%) |
|---|---|---|---|---|
| EfficientNetB0 | 97.43 ± 0.98 | 97.33 ± 0.80 | 97.06 ± 0.78 | 97.11 ± 0.83 |
| MobileNetV1 | 97.62 ± 0.85 | 97.59 ± 1.01 | 97.32 ± 0.89 | 97.37 ± 0.97 |
| MobileNetV2 | 97.43 ± 1.72 | 97.37 ± 1.86 | 97.09 ± 1.60 | 97.20 ± 1.73 |

Average accuracy, precision, recall, F1-score, and ± standard deviations were calculated from 5-fold Cross-Validation.

## 5.2. Performance of Lite-FBCN Models

We evaluated the effectiveness of the proposed Lite-FBCN model in brain disease classification tasks, employing the same training and evaluation procedures as the baseline models. Table 2 presents the performance metrics of the Lite-FBCN model. The results show that Lite-FBCN models achieved higher accuracy compared to baseline models when using the MobileNets backbone. Conversely, there was a decrease in performance for EfficientNetB0, though we found no statistically significant difference between both methods (P = 0.5659). The choice of backbone architecture significantly influences the performance of Lite-FBCN (P = 0.0047). However, further analysis suggests that differences in performance between MobileNetV1 and MobileNetV2 may not be significant (P = 1). This implies that EfficientNetB0 exhibited the lowest performance when utilized as a backbone in Lite-FBCN models. Meanwhile, the CRF ($\gamma$) also did not play a significant role in performance (P = 0.6462).

**Table 2.** Performance of Lite-FBCN on the 5-fold cross-validation data

| $\gamma$ | Backbone | Acc (%) | Prec (%) | Rec (%) | F1-score (%) |
|---|---|---|---|---|---|
| 2 | EfficientNetB0 | 91.05 ± 2.61 | 91.38 ± 2.36 | 91.05 ± 2.61 | 90.64 ± 2.92 |
|   | MobileNetV1 | 98.00 ± 0.92 | 98.04 ± 0.94 | 98.00 ± 0.92 | 97.99 ± 0.93 |
|   | MobileNetV2 | 98.29 ± 0.49 | 98.33 ± 0.48 | 98.28 ± 0.49 | 98.18 ± 0.56 |
| 4 | EfficientNetB0 | 93.81 ± 1.62 | 93.79 ± 1.66 | 93.81 ± 1.62 | 93.61 ± 1.67 |
|   | MobileNetV1 | 98.10 ± 0.90 | 98.21 ± 0.88 | 98.10 ± 0.90 | 98.09 ± 0.91 |
|   | MobileNetV2 | 97.81 ± 0.49 | 97.89 ± 0.44 | 97.81 ± 0.49 | 97.77 ± 0.50 |
| 8 | EfficientNetB0 | 93.33 ± 2.48 | 93.49 ± 2.47 | 93.33 ± 2.48 | 93.20 ± 2.58 |
|   | MobileNetV1 | 97.52 ± 0.76 | 97.59 ± 0.76 | 97.32 ± 0.57 | 97.52 ± 0.76 |
|   | MobileNetV2 | 97.52 ± 0.70 | 97.62 ± 0.67 | 97.52 ± 0.70 | 97.50 ± 0.70 |

Average accuracy, precision, recall, F1-score, and ± standard deviations were calculated from 5-fold Cross-Validation.

## 5.3 Comparison with Existing Bilinear Methods

To further validate the performance of the Lite-FBCN model, we assessed the performance of existing bilinear CNN models, including both standard BCNNs and Fast BCNNs. These comparisons were based on the same dataset and evaluation metrics. Due to the observed poor performance of EfficientNetB0 backbone in the Lite-FBCN model, we focused the comparison only on MobileNetV1 and MobileNetV2 backbones. Table 3 summarizes the performance results of the existing bilinear CNN and Fast BCNN models. Statistical analysis showed no significant performance difference between implementing the existing bilinear methods and Lite-FBCN models ($P$ = 0.1759). No significant difference was also found between conventional Fast BCNNs and Lite-FBCN models with MobileNetV1 as backbone ($P$ = 0.4572). However, a statistically significant difference was observed when comparing both methods using MobileNetV2 as backbone ($P$ = 0.0087), suggesting that channel reducer layer in Lite-FBCN model can help maintain performance using this backbone. Ultimately, the choice of method should also consider computational efficiency.

Table 3. Performance of the existing conventional Bilinear CNNs and Fast BCNNs on the 5-fold cross-validation data.

| Pooling Method | Backbone | Acc (%) | Prec (%) | Rec (%) | F1-score (%) |
|---|---|---|---|---|---|
| BCNNs | MobileNetV1+ MobileNetV2 | 98.19 ± 3.00 | 98.27±17.28 | 98.19 ± 5.68 | 98.19 ± 5.68 |
| Fast BCNNs | MobileNetV1 | 97.62 ± 3.00 | 97.73 ± 2.67 | 97.62 ± 7.71 | 97.63 ± 7.71 |
| | MobileNetV2 | 96.95 ± 3.00 | 97.04±17.00 | 96.95 ± 5.68 | 96.91 ± 5.68 |

Average accuracy, precision, recall, F1-score, and ± standard deviations were calculated from 5-fold Cross-Validation.

**5.4 Comparison on Computation Efficiency**

In addition to classification performance, we evaluated the computational efficiency of the Lite-FBCN model. This analysis involved measuring the inference time and the number of parameters during model deployment. We compared these metrics against those of baseline CNNs and existing bilinear models. The findings, presented in Table 4, show the inference time and parameter count comparison between different bilinear models with various CNN backbones. Average inference time and standard deviations were calculated from 5 repeated measurements.

Table 4. Comparison of inference time and parameter count between various bilinear models using different CNN backbones.

| Methods | $\gamma$ | Backbone | #Params (million) | Inference time (ms/img) |
|---|---|---|---|---|
| **Baseline** | N/A | EfficientNetB0 | 4.8381 | 43.48 ± 1.19 |
| | | MobileNetV1 | 3.8863 | 14.50 ± 0.44 |
| | | MobileNetV2 | 3.0465 | 24.94 ± 1.27 |
| **BCNN** | N/A | MobileNetV1 + MobileNetV2 | 17.2833 | 45.54 ± 2.92 |
| **FBCNN** | N/A | MobileNetV1 | 12.6661 | 16.39 ± 0.54 |
| | N/A | MobileNetV2 | 17.0036 | 25.94 ± 1.27 |
| **Lite-FBCN** | 2 | EfficientNetB0 | 8.5558 | 45.28 ± 1.35 |
| | | MobileNetV1 | 7.5713 | 15.89 ± 0.65 |
| | | MobileNetV2 | 6.7642 | 29.72 ± 2.69 |
| | 4 | EfficientNetB0 | 5.3811 | 46.64 ± 1.52 |
| | | MobileNetV1 | 4.4785 | 15.29 ± 0.56 |
| | | MobileNetV2 | 3.5895 | 28.92 ± 1.65 |
| | 8 | EfficientNetB0 | 4.4849 | 45.05 ± 0.39 |
| | | MobileNetV1 | 3.6232 | 16.56 ± 0.51 |
| | | MobileNetV2 | 2.6933 | 29.73 ± 2.47 |

From these results, significant reductions in computational costs and inference time are achieved by the Lite-FBCN model, particularly with higher

CRF (γ). While the lightest parameter count is attained with γ = 8, using γ = 4 results in faster inference times across all backbones. Among the backbone networks, MobileNetV1 stands out for its impressive inference time, outperforming other models in acceleration. Notably, MobileNetV1 demonstrated the fastest inference time with γ = 4.

**5.5 Further Evaluation on Subject-wise Classification**

In this section, we present an additional evaluation of the proposed Lite-FBCN model and existing bilinear models using the hold-out dataset. This evaluation provides further insights into the model's performance and robustness in a real-world scenario, outside the cross-validation framework. The analysis focuses on the model's ability to correctly classify brain diseases across different new subjects, ensuring that the performance improvements were not due to overfitting to specific individuals. For this evaluation, we collected new brain MR images of axial T2-weighted scans, categorized similarly to the cross-validation data. Specifically, we obtained 1091 images of CN from 25 subjects from the IXI dataset. Additionally, we acquired 3559 images of NEO from 167 subjects from the BraTS dataset, 102 images of CVA from the Kurzad Poyraz dataset [28], 716 images from 65 subjects from the ADNI dataset, and 287 images from 35 subjects from Radiopaedia [37], [38].

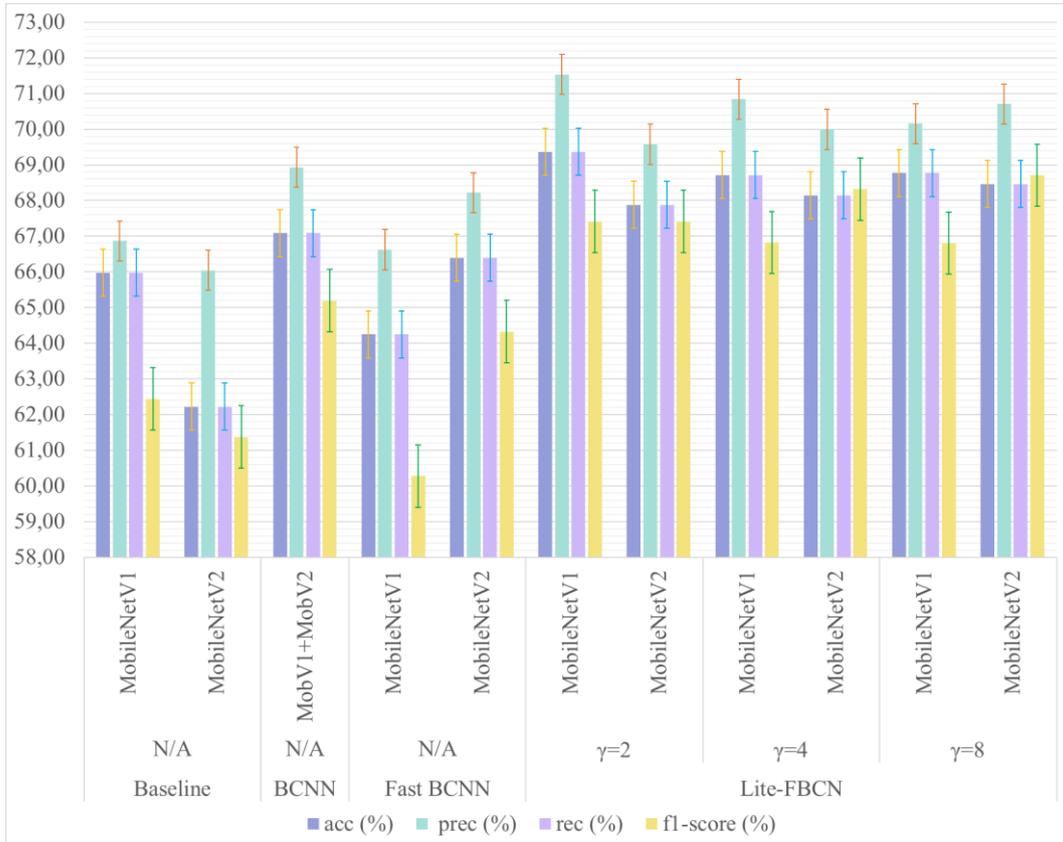

**Figure 3.** Performance comparison of the proposed Lite-FBCN and existing bilinear models on the hold-out data.

The results of the evaluation on hold-out data are shown in Figure 3. We observed that Lite-FBCN with MobileNetV1 as the backbone network performed the best compared to other models, particularly when γ = 2. Even with higher γ values, Lite-FBCN with MobileNetV1 continued to outperform other backbones and models. There was no statistically significant difference in Lite-FBCN performance for MobileNetV1 across different γ values (P = 0.7142). Notably, while MobileNetV1 with Lite-FBCN showed the best performance, there was no significant difference between using MobileNetV1 or MobileNetV2 as the backbone (P = 0.8876). Additionally, there was no significant performance difference between Lite-FBCN models and both baseline Fast BCNNs (P = 0.2579) and BCNNs (P = 0.5467).

The confusion matrix comparison in Figure 4 provides a detailed breakdown of classification accuracy across different classes, underscoring the robustness of the Lite-FBCN model, particularly in correctly identifying various brain disease categories compared to the baseline and existing bilinear models. While all models struggled with classifying NDD and INF, especially where these classes were often confused with other closely similar-looking classes, Lite-FBCN achieved better accuracy for these classes compared to the others. The choice of backbone networks between MobileNetV1 and MobileNetV2 also appears to influence accuracy for each class, although not as significant as the implementation of bilinear methods.

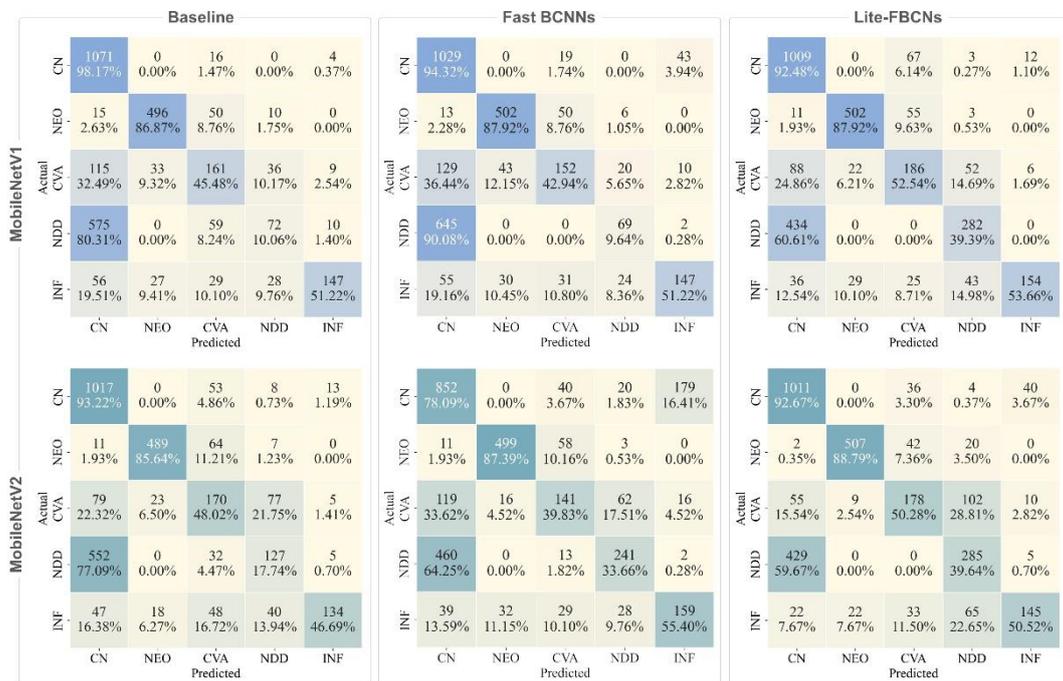

**Figure 5.** Confusion matrix comparison of the proposed Lite-FBCN and existing bilinear models on the hold-out data.

The UMAP visualization in Figure 5 provides further insights into model performance by illustrating the clustering and separation of different brain disease categories achieved by the various models. This visualization highlights the superior feature representation capabilities of the Lite-FBCN model. While baseline models exhibit more scattered clusters, the implementation of bilinear pooling methods results in improved clustering, despite some overlapping between closely related brain classes. Notably, classification becomes more distinct in bilinear models, particularly in Lite-FBCN. Similar to the results observed in the confusion matrix, the choice of backbone networks also appears to affect clustering in UMAP, although not as significantly as the implementation of bilinear pooling methods.

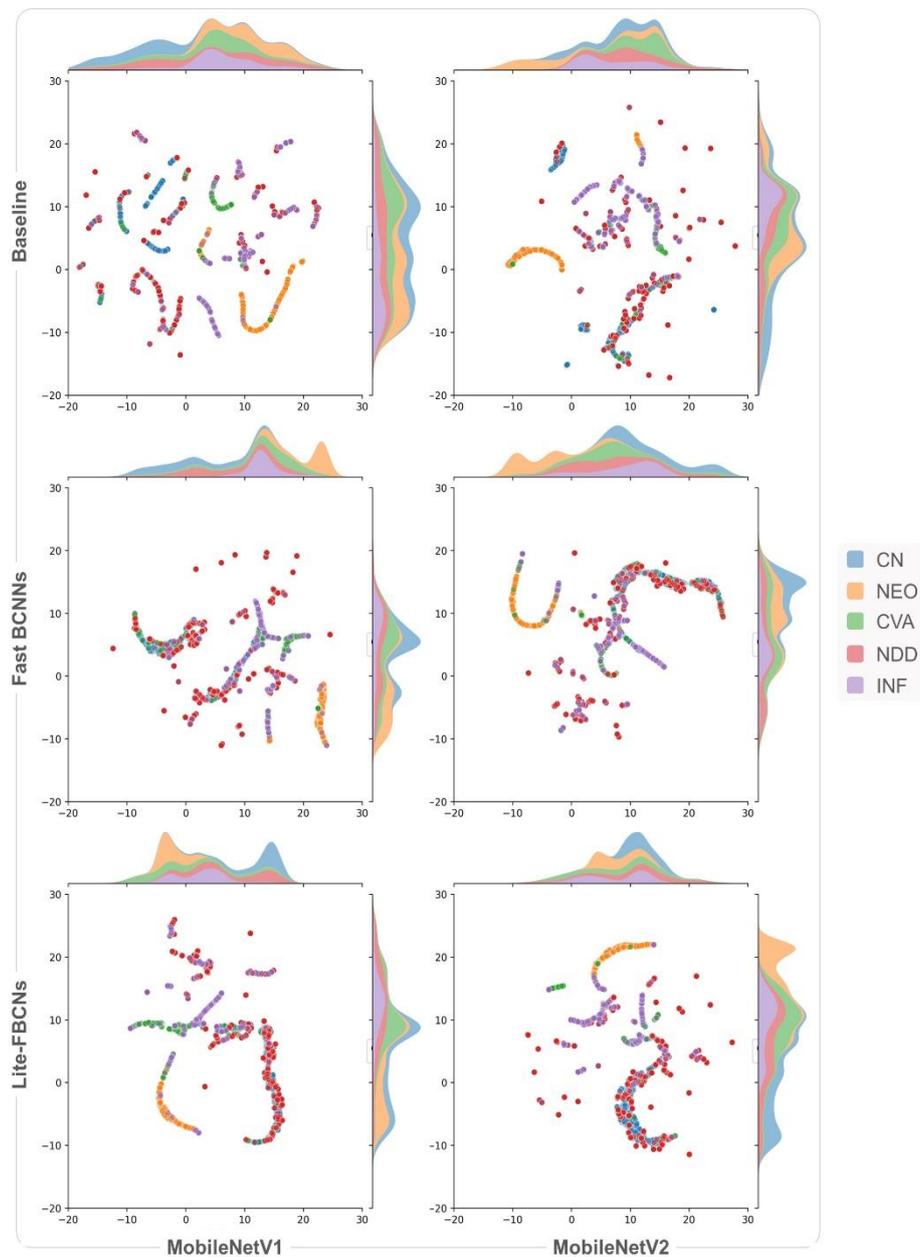

**Figure 4.** UMAP visualization comparison of the proposed Lite-FBCN and existing bilinear models on the hold-out data.

**5.6 Discussion**

The Lite-FBCN model outperforms both baseline models and existing bilinear models in brain disease classification tasks. Extensive evaluations on both cross-validation and hold-out datasets consistently demonstrate that Lite-FBCN excels, even though the superiority is not statistically significant.

Within each method, the choice of backbone networks did not have a significant impact. However, when MobileNets were used as backbones in bilinear methods, they demonstrated better performance compared to baseline models, particularly in Lite-FBCN. Although it is worth noting that no statistically significant difference was found during cross-validation and hold-out data evaluation. Conversely, EfficientNetB0, when used as the Lite-FBCN backbone, showed lower performance and was more computationally expensive. The discrepancy may be due to its complexity and deeper network architecture, making it less suitable for integration as a bilinear backbone. EfficientNetB0's intricate design might lead to overfitting or excessive computational overhead, reducing its effectiveness as a bilinear backbone.

The evaluation also revealed that the channel reduction factor (CRF) did not significantly impact classification performance of Lite-FBCN model. It appears that the CRF has minimal impact on the model's feature extraction and utilization for brain disease classification, despite compressing the feature map dimensionality before bilinear pooling. The robust and efficient feature extraction of the Lite-FBCN architecture minimizes dependency on the exact dimensionality of the feature maps, resulting in a concise design. Lite-FBCN achieves a balance between feature richness and model simplicity by using a single lightweight CNN backbone and a channel reducer. This allows it to preserve critical information across different CRF values. In addition, bilinear pooling efficiently captures higher-order feature interactions, improving the representation of intricate patterns and subtle variations in MRI images, ensuring essential features are properly captured regardless of the initial channel reduction.

The consistent performance of Lite-FBCN, unaffected by the CRF, demonstrates the model's strong architecture and effective feature extraction capabilities. However, MobileNetV1 with a CRF of 4 appears to be the most recommended configuration based on the combined factors of computational efficiency and robustness in classification performance. A CRF of 4 is lighter compared to a CRF of 8 because it reduces the dimensionality of feature maps to a lesser extent, which leads to a smaller reduction in the number of channels. This results in a lower computational load and faster processing times while maintaining a sufficient level of feature richness. A CRF of 8 would compress the feature maps more aggressively, potentially leading to loss of critical information and increased computational demands.

We assessed the models by analyzing their confusion matrices and UMAP visualizations during hold-out data evaluation. The results show that distinguishing between NDD and INF classes posed a challenge for all models, with NDD often being confused with CN, likely due to their similar appearance or shared features. However, Lite-FBCN showed significant improvements in accurately distinguishing these classes. Additionally, the UMAP visualization further underscored the superior feature representation capabilities of Lite-FBCN, with clearer clustering and separation of disease categories compared to baseline and existing bilinear models.

Lite-FBCN owes its success to its innovative design, which includes a single-network architecture with transfer learning for feature extraction and the addition of a channel reducer before the bilinear pooling layer. This design choice leads to a more compact bilinear vector and a smaller model size compared to existing bilinear methods, reducing computational overhead and enhancing the model's ability to capture complex feature interactions crucial for accurate disease recognition. Further research can investigate different backbone architectures and optimization techniques to improve the efficiency and performance of Lite-FBCN. This could expand its potential use in various clinical settings, ensuring suitability for deployment in resource-constrained clinical environments and real-time applications.

## 6. CONCLUSION

After conducting thorough research and evaluation, it is evident that the Lite-FBCN model outperforms both baseline CNN models and existing bilinear CNN models in classifying brain diseases. Lite-FBCN exhibits superior performance in various brain disease categories, consistently achieving higher accuracy, precision, recall, and F1-score. Its robustness and ability to generalize are clearly demonstrated in the evaluation of hold-out datasets. The model's efficient computational balance and manageable complexity make it well-suited for real-time applications and clinical environments with limited resources. Backbone networks like MobileNetV1 and MobileNetV2, along with bilinear pooling methods, have been found to greatly improve feature representation and classification accuracy. This has been observed through visualization techniques such as UMAP and confusion matrix analysis. The results highlight the potential of the proposed Lite-FBCN for use in clinical environments with limited resources and for real-time applications. Future research should focus on improving performance and seamlessly integrating it into diagnostic workflows.